# Ferroelectric MirrorBit-Integrated Field-Programmable Memory Array for TCAM, Storage, and In-Memory Computing Applications

*Paritosh Meihar[1], Rowtu Srinu[1], Sandip Lashkare[1], Ajay Kumar Singh[1], Halid Mulaosmanovic[2], Veeresh Deshpande[1], Stefan Dünkel[2], Sven Beyer[2] and Udayan Ganguly[1]*

[1] Department of Electrical Engineering, Indian Institute of Technology Bombay, Mumbai, 400076, India
[2] GlobalFoundries Fab1 LLC and Co. KG, 01109 Dresden, Germany

**ABSTRACT**

In-memory computing on a reconfigurable architecture is the emerging field which performs an application-based resource allocation for computational efficiency and energy optimization. In this work, we propose a Ferroelectric MirrorBit-integrated field-programmable reconfigurable memory. We show the conventional 1-Bit FeFET, the MirrorBit, and MirrorBit-based Ternary Content-addressable memory (MCAM or MirrorBit-based TCAM) within the same field-programmable array. Apart from the conventional uniform Up and Down polarization states, the additional states in the MirrorBit are programmed by applying a non-uniform electric field along the transverse direction, which produces a gradient in the polarization and the conduction band energy. This creates two additional states, thereby, creating a total of 4 states or 2-bit of information. The gradient in the conduction band resembles a Schottky barrier (Schottky diode), whose orientation can be configured by applying an appropriate field. The TCAM operation is demonstrated using the MirrorBit-based diode on the reconfigurable array. The reconfigurable array architecture can switch from AND-type to NOR-type and vice-versa. The AND-type array is appropriate for programming the conventional bit and the MirrorBit. The MirrorBit-based Schottky diode in the NOR-array resembles a crossbar structure, which is appropriate for diode-based CAM operation. Our proposed memory system can enable fast write via 1-bit FeFET, the dense data storage capability by Mirror-bit technology and the fast search capability of the MCAM. Further, the dual configurability enables power, area and speed optimization making the reconfigurable Fe-Mirrorbit memory a compelling solution for In-memory and associative computing.

## 1. INTRODUCTION

Von Neumann architecture has been the dominant governing design for the digital computing systems for over 7 decades [1]. As a result of the transistor scaling (predicted by Moore's law), the processing speed continued to increase and reached a point where the data transfer rate from the memory to the processor by the system bus became the limiting factor and thereby increased the latency and the communication cost. This is known as "Von Neumann Bottleneck" [2]. The broadening of the gap between the processor and memory speed, which is also known as the "memory wall" [3], and the explosive growth of the data due to rise in the machine learning applications, known as "Big-data", have further rendered the Von Neumann architecture inefficient for the data-intensive tasks [2, 4].

"In-memory computing (IMC)", an alternate concept, tackles this problem by eliminating the need for the data movement which consumes energy and increases latency [5, 6]. The conventional memory unit in the Von-Neumann architecture is replaced with a memory array which performs certain computational tasks within itself and therefore termed as "Computational memory array". The computation in IMC is accomplished by utilising the memory device properties, the array structure, and the peripheral circuit including control logic [7].

Content-addressable memory (CAM) [8], in particular, is a special form of in-memory computing circuit used in various applications like high speed search [9], pattern recognition [10], and deep neural networks [11], etc. The SRAM-based CAMs posed some practical implementation challenges such as, volatility, huge area overhead (per cell 16 transistors) [12] and huge power consumption [13]. This shifted the focus toward emerging non-volatile memories [14], including, resistive RAM (RRAM), magnetic tunnel junction (MTJ), ferroelectric capacitor (FeCAP)/ferroelectric field-effect transistor (FeFET) [15]. After the discovery of ferroelectricity in $HfO_2$ in 2011 [16], the research on the ferroelectric devices resumed at an accelerated rate. Ferroelectric devices offer various advantages, such as high scalability, low write power, fast switching, and CMOS compatibility [17]. Due to such remarkable properties, the ferroelectric devices have also been explored for the CAM applications [18, 19].

Although, IMC using emerging memories show an attractive promise in improving computation efficiency, the applications of these architectures are restricted to a limited set of domain-specific computational kernels, such as pattern matching [20] and matrix multiplication [21]. As growing machine learning-based data-driven applications, the lack of programmable and flexible memory array makes it difficult to adapt to the future technological changes in the workload.

In this report, we present a fully reconfigurable ferroelectric memory cell and a reconfigurable array architecture (Fig. 1 (a)), which can be morphed into the required memory type along with corresponding architecture for in-memory computation as well as the storage. There are at least three memory types that can be configured by applying appropriate field, a) the conventional 1-bit fast (low density) memory, b) the Ferroelectric MirrorBit-based high density memory, and c) the reconfigurable MirrorBit diode-based Ternary CAM (MCAM), are shown (Fig. 1 (c)). This field-programmable memory array (FPMA), that can switch between AND-type and NOR-type array (Fig. 1 (b)), will enable application-based optimization of the memory allocation for higher energy efficiency and performance. Here, we show the programming of the MirrorBit and the MCAM operation. The programming of the conventional FeFET has not been presented separately, as it is contained in the programming of the MirrorBit and it has also been shown in our previous work [22], as well as, reported in the literature [23].

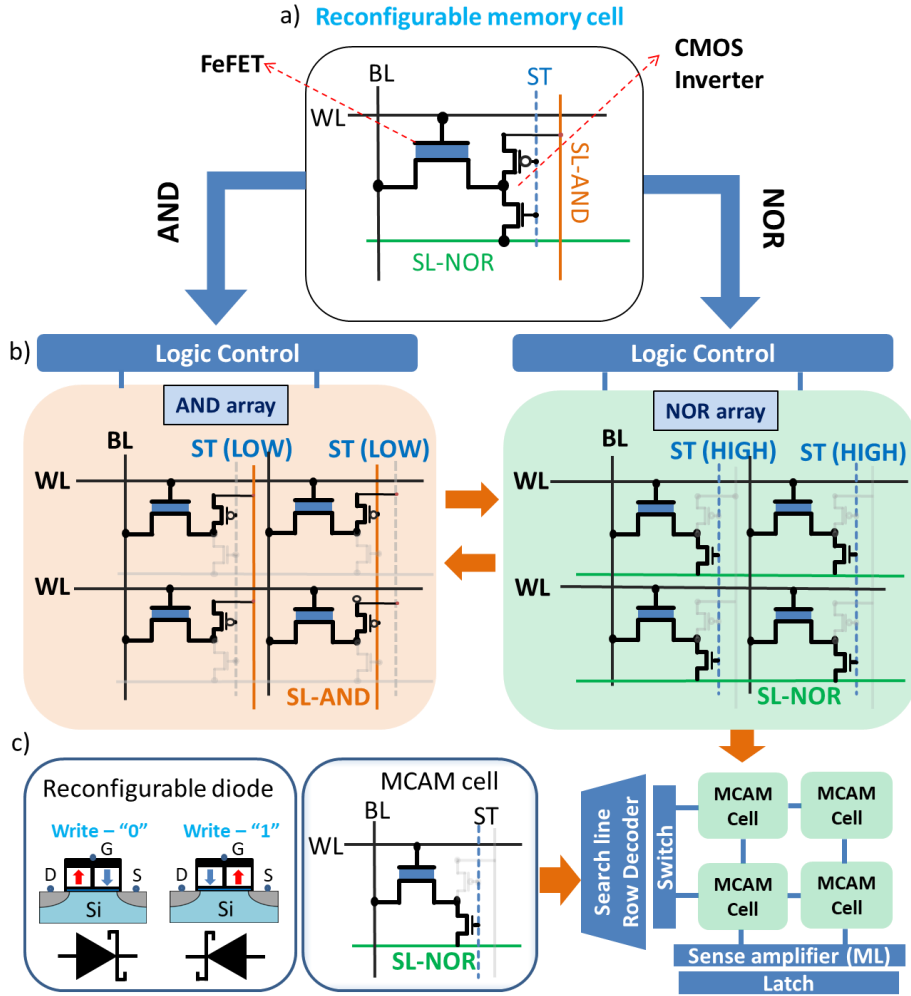

**Figure 1**. **Reconfigurable memory cell:** a) and b) The reconfigurable memory cell consists of 1-FeFET and 1-inverter. The inverter's one end is connected to an AND source line (SL-AND) and another is connected to the NOR source line (SL-NOR). By HIGH and LOW at Select transistor (ST), the array type can be changed to AND and NOR, respectively, c) The MirrorBit states Source Write (SW) and Drain Write (DW) are modelled as a Schottky diode. The MirrorBit-based CAM cell (MCAM) on a NOR architecture and MCAM array along with peripherals are shown.

## 2. METHODS

The memory cell consists of 1 FeFET and 1 inverter. The FeFET is fabricated at GlobalFoundries' (GF) 28 nm bulk-HKMG CMOS technology [24]. The dimensions of the device used for the measurements are $L = 240\ nm$ and $W = 240\ nm$. The inverter used for the measurements, is 'MM 74C04N' IC. The gate of the FeFET is connected to the Wordline (WL) of the array, the drain is connected to the Bitline (BL), and the source is connected to the output pin of the inverter (Fig. 1). The input pin of the inverter is connected to select transistor (ST) line. The $V_{CC}$ of the inverter IC is connected to the Sourceline-AND (SL-AND), i.e. this line becomes source line in the AND configuration when the input pin is LOW. Similarly, the GND pin is connected to the Sourceline-NOR (SL-NOR). Depending on the bias on the ST, only one of the array type is active during an operation. The electrical characterization is performed using the Agilent B1530A 4-channel Waveform generator/Fast measurement unit (WGFMU). The 4 channels of the WGFMUs and GND are connected to WL, BL, $V_{CC}$, Input pin, and GND pin respectively.

A TCAD model is also developed and calibrated using GF-FeFET devices [22] to explain the working of the MirrorBit and the origin of the reconfigurable diode-like behaviour. A calibrated Spice model is presented to simulate the MCAM operation on the reconfigurable array.

## 3. RESULTS AND DISCUSSION

### A. Field-programmable memory array (FPMA)

The capability of adapting to application-based architectures has become the driving factor for the reconfigurable architectures. This provides the system an ability to optimize to gain higher performance per Watt over variety of applications [25]. The proposed reconfigurable architecture (Fig. 1 (a) and (b)), which is termed as "Field-Programmable Memory Array (FPMA)", can transform from AND-type to NOR-type and vice-versa by applying LOW and HIGH voltage at the ST line. The AND-type seems to be the appropriate array for the conventional 1-Bit and the MirrorBit [26]. However, the MCAM operation requires a NOR-type array (also known as crossbar array for two terminal devices). The array reconfigure operation can be performed in parallel to the programming operation and hence does not add any additional latency.

The FPMA enables us to reallocate resources for the optimum performance, which is also referred to as "Adaptive resource partitioning". For instance, during neural network training, fast 1-Bit memory can be configured and along with it a part of the array can be programmed as TCAM for faster search operations. During the later stage of training or during inference, the storage can be increased through the MirrorBit.

### B. Conventional-bit and MirrorBit in the FPMA

In our previous work [22], the FeFET-based MirrorBit operation has been presented. The conventional FeFET has a uniform polarization switching along the channel when the write voltage is applied at the gate (keeping all the other terminals grounded) giving rise to 2 states, Uniform Low $V_T$ (UWL) and Uniform High $V_T$ (UWH) (Fig. 2(a) and (b)). The threshold voltages are obtained through the transfer characteristics measured at $V_D = 0.1\ V$ [22].

Starting with UWL state, when a $4\ V$, $100\ \mu s$ pulse is applied at the SL-AND keeping all other terminals grounded, a non-uniform electric field is produced causing a switching in the polarization near the source in a gradient fashion, which is referred to as Source Write (SW) state (Fig. 2 (c)). Similarly, the drain write (DW) state is programmed (Fig. 2(d)), by applying a HIGH voltage at the ST and keeping the other terminals grounded. To read the MirrorBit states, a read voltage of $1.5\ V$ is applied at SL-AND along with sweeping gate voltage from $V_G = -0.5\ V$ to 1.5 $V$, referred to as Drain Read (DR), and similarly a read voltage at the BL keeping SL-NOR grounded is referred to as source read (SR) (Fig. 2 (e) and (f)). As we can observe, for the UWL and UWH states, both SR/DR $V_T$s are low and high respectively. However, for the DW state, the SR $V_T$ is lower but the DR $V_T$ is relatively higher. Similarly, for SW state, the SR $V_T$ is higher but the DR $V_T$ is relatively lower. The measured transfer curves and corresponding $V_T$s of all 4 states are summarized in Fig. 2(g) and (h).

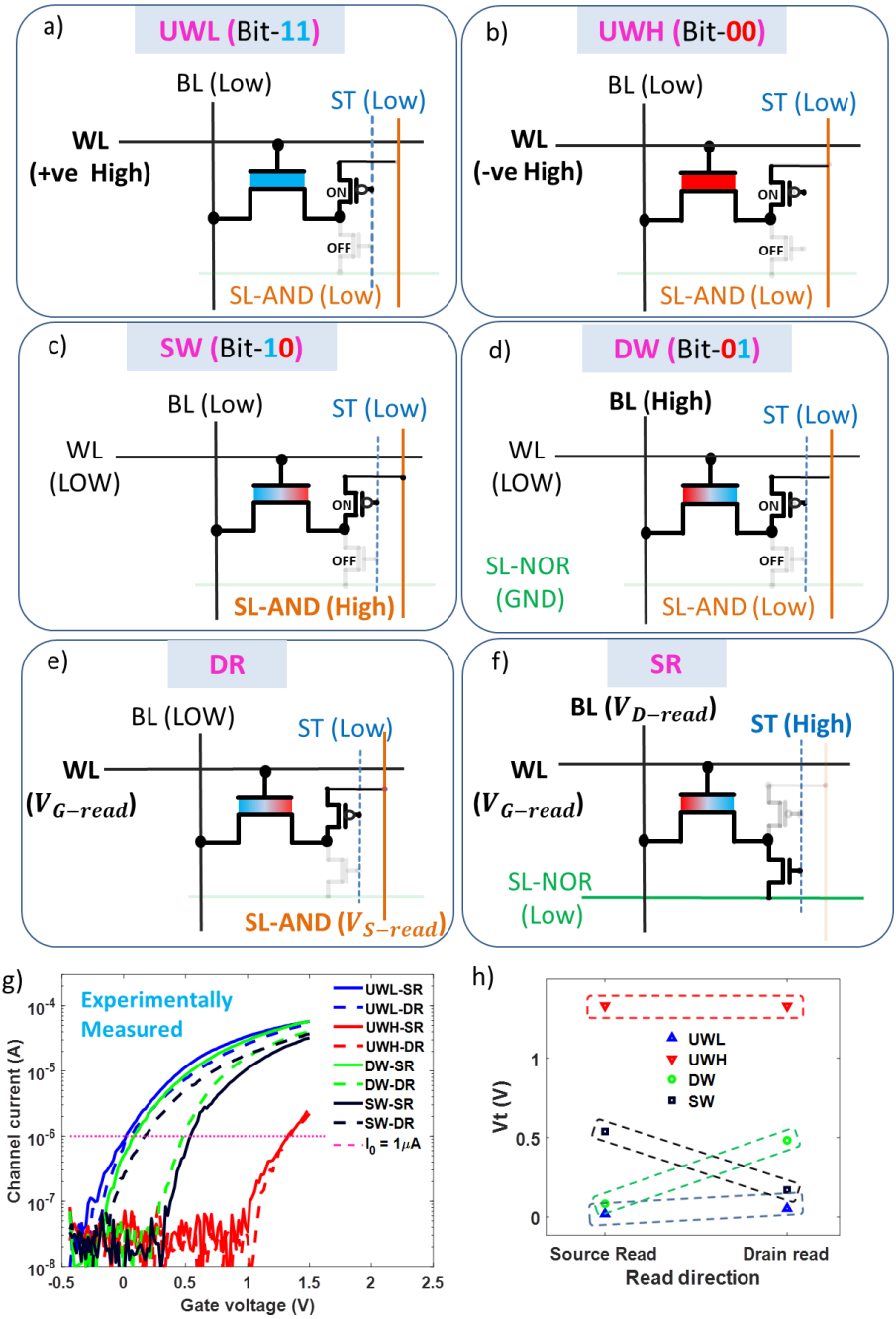

**Figure 2. MirrorBit write and read**: a), b), c), and d) The programming of uniform write high $V_T$ (UWH), uniform write low $V_T$ (UWL), Source Write (SW), and Drain Write (DW) states using the appropriate Select transistor line voltage. e) and f) To read the MirrorBit states, we measure the channel current in both directions ($V_{read} = 1.5\ V$ at Drain/Source), namely, Drain Read (DR) and Source Read (SR). g) and h) Experimentally measured transfer curves and corresponding $V_T$s for all the four states.

The TCAD model reveals that the conduction band shape in the case of DW/SW resembles a Schottky barrier (Fig. 3 (a) and (b)) [22]. The DR and SR are experimentally measured for the SW state in the NOR-type configuration. In the output curve of the SW state (Fig. 3 (d)), we see that for drain read voltage varying from $0\,V$ to $1.5\,V$, keeping $V_G = 0.2\,V$, $V_S = 0\,V$, and $V_{sub} = 0\,V$, gives the ON region of the MirrorBit diode. Similarly, varying the source voltage from 0 V to 1.5 V gives the OFF region. The current characteristics appear to be similar to that of a Schottky diode. There is an excellent match between the experimental data (blue) and the TCAD model (green), shown in Fig. 3(d). Using this current characteristic, the diode in the Spice model is also calibrated. In the operation region ($V_{read} = 1.5\,V$), the $R_{ON}$ and $R_{OFF}$ of the diode is set appropriately to match the experimental current levels.

### C. MirrorBit-based Ternary Content-Addressable Memory (MCAM)

In the conventional 2-FeFET TCAM cell, a constant matching line (ML) voltage is considered as a match and different discharge rates correspond to the degree of mismatch [19]. Here, we utilize the directional conductive property of the diode and reconfigure the diode orientation appropriately to write the CAM states. The Write "0" state and Write "1" states are shown in Fig. 4. The search voltages ($V_{read} = 1.5\,V$) at Search Line 1 (SL1) and Search Line 2 (SL2/ML) are appropriately applied such that diode remains in OFF region giving $I_{OFF}$ at the ML for a match and $I_{ON}$ for a mismatch. There is another state, required in some applications, called the "don't care" state (Write "X"). The Write "X" is the UWH state, which gives OFF current for the reads, i.e. a match for both the reads. This state is always considered as a match. All the possible combinations of Write states and read conditions along with matching criteria are summarized in Fig. 4. The SL1/SL and SL2/ML are interchangeably used later in this report for simplicity.

A 5 x 5 MCAM array is simulated in Spice (Fig. 5). The data in the MCAM cells are written through the DW and SW programming, following which the array is reconfigured to NOR-type by applying logic "1" at the ST. The NOR array resembles a crossbar structure, which is appropriate for performing MCAM search [27]. In the diode based crossbar array, the entire search string cannot be searched in one shot.

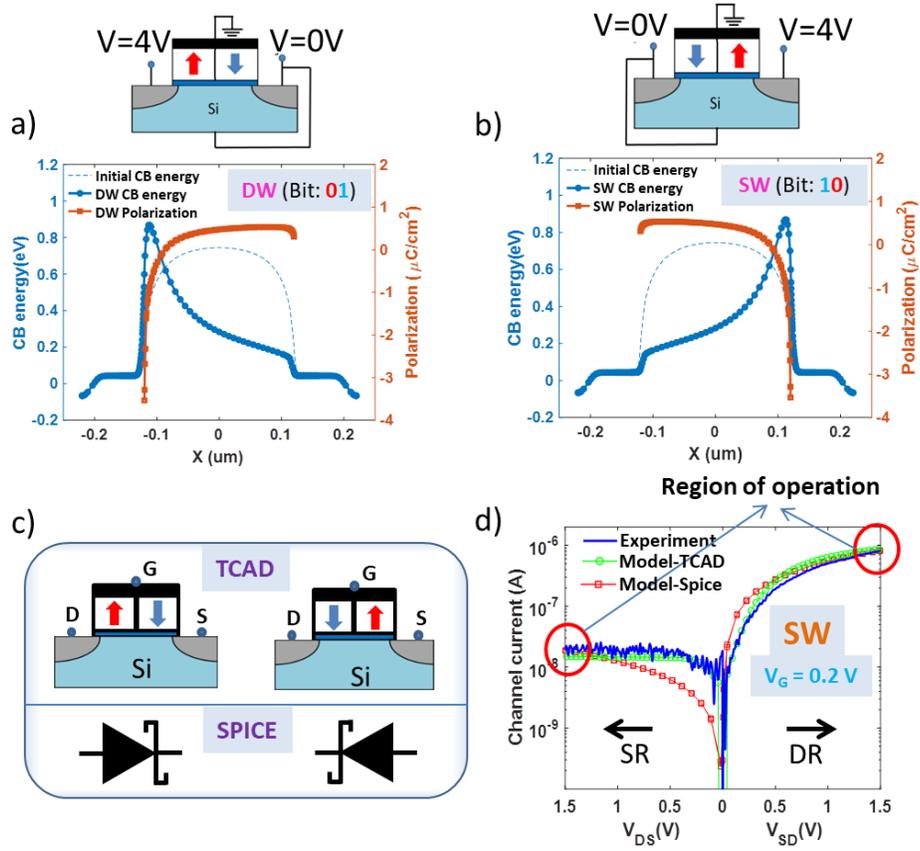

**Figure 3. MirrorBit based reconfigurable diode:** a) and b) the schematic of the device during the write operation and after the biases are removed the polarization/conduction band variation of the SW and DW states are plotted. The TCAD model reveals that the variation in the conduction band energy resembles a Schottky diode. c) The compact modelling of the DW and SW in the Spice. d) Experimentally obtained current characteristics of the SW state. There is a Schottky diode-like resemblance and an excellent match with the characteristics obtained from the TCAD model. The characteristics are further used to calibrate a Spice model in the operating regions to simulate the MCAM operation in Spice.

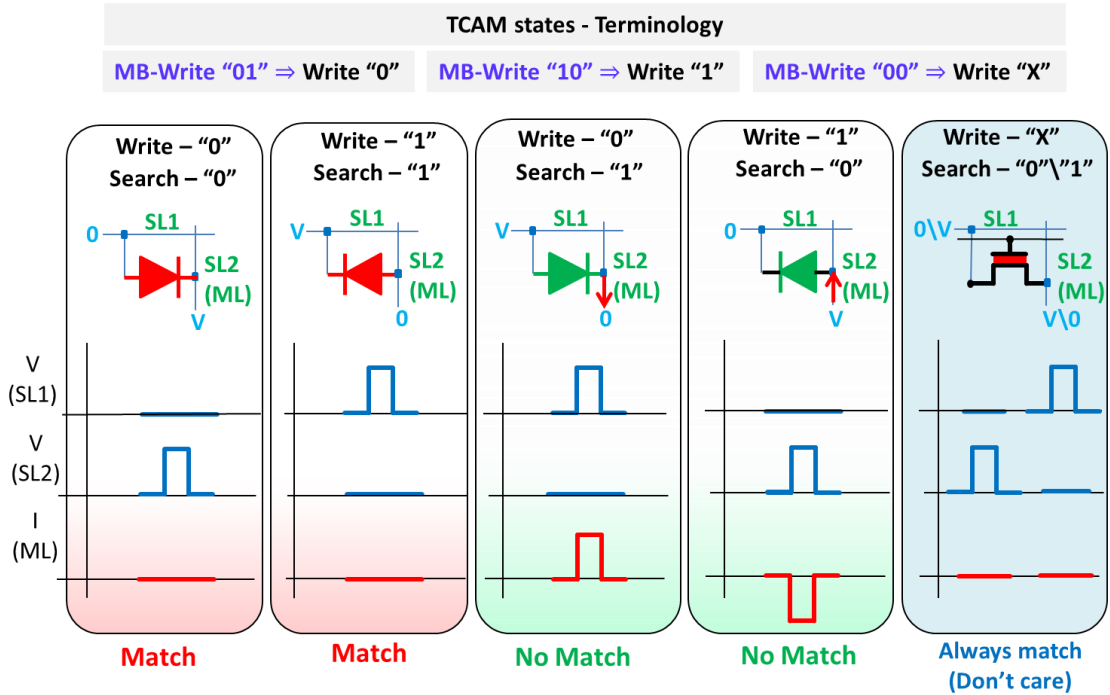

**Figure 4. MCAM pulse scheme:** There are three write states in the MCAM, Write "1", Write "0" and Write "X" and two search operations search "1" and search "0". In case of a match (Write "0" and search "0", the diode goes in the OFF region. In case of a mismatch, the diode turns ON. For a don't care ("X") condition the state UWH is programmed, which gives OFF current for both searches.

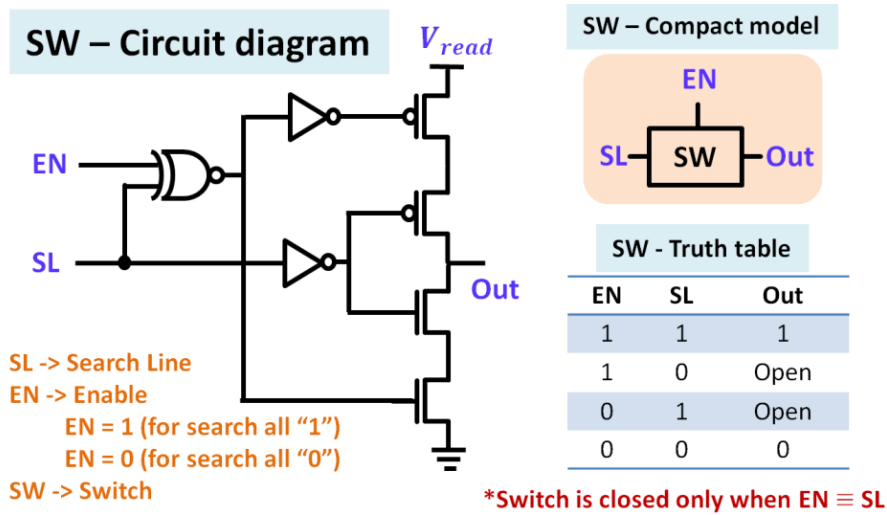

**Figure 5. Search criteria:** Switch (SW) circuit has two inverters, a XNOR gate, and a three-stage logic to implement a two-step read. EN is "1" for the search all "1"s operation, and "0" for search all "0"s. When EN = 1 (search all "1"s), the SW corresponding to the SL rows with search "1" bits are closed and the $V_{read}$ appears at the Out pin (the remaining SWs are open). Similarly, for EN = 0, the SWs for search "0" bits are closed, and Out pin gets grounded. The truth table summarizes all the search criteria

The search "1" row current can split and flow into the search "0" rows instead of going into the ML as they both are at the same voltage. This would give the false current levels at the ML. To avoid such crosstalk, two-step read method is proposed and implemented. The read operation consists of two parts, search all "1"s and search all "0"s. A special "switch circuit" (SW) is designed to achieve the two-step read (Fig. 5). In the search all "1"s step, the SW will close for the rows where the SL is "1" and remain open for remaining rows, where, the SL is "0". Similarly, in the search all "0"s, the SW will close for the rows, where SL is "0".

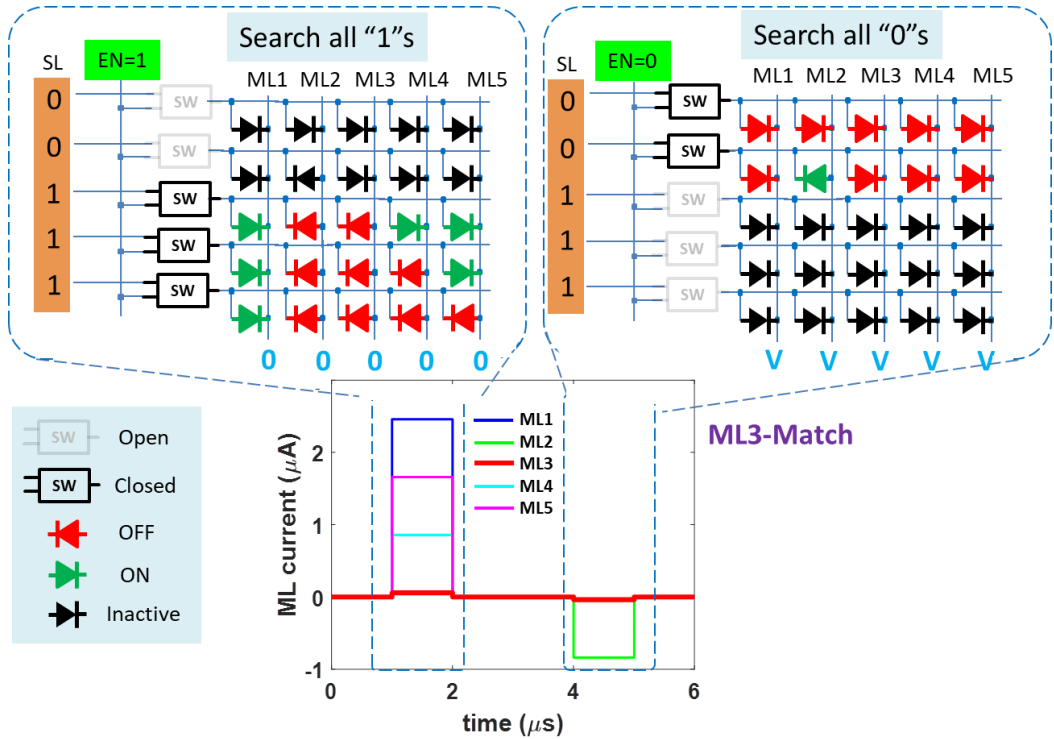

**Figure 6. MCAM operation:** After the diodes are programmed in the array, the search operation is divided into two parts, Search all "1"s and Search all ``"0"s. During the search "1", the search "0" lines are open and the Matching lines (MLs) are grounded. During the search "0", the search "1" lines are open and MLs are at the read voltage. The ML with OFF current for both the reads represents a match

In the search "1" operation, after the SW is closed for search "1" input (other SL rows are open), a read voltage pulse ($V_{read} = 1.5\ V$) appears at the "out" node of the SW, which is the input for the MCAM array (Fig. 6). The ML columns are kept at $0\ V$ (the sense amplifier's preset voltage is set to $0\ V$). The forward and reverse bias diodes give a cumulative current at the MLs. The sense amplifiers output is stored in a Latch. Similarly, in the search "0" operation, the SW for the SL rows with "0" bit search are closed (other rows' SWs are open). The ML columns, in this case, are kept at $1.5\ V$ (the sense amplifier's preset voltage is set to $1.5\ V$). The cumulative current at the MLs are sensed and stored in the Latch. The latch corresponding to ML with the lowest current for both searches is the exact match for the corresponding searched vector. Both searches can be performed in a single pulse cycle, i.e. search "1" operation, when the pulse is high and search "0", when the pulse is low. Here, the latch-type sense operation is omitted, as this is a standard technique [28]. The diodes in the black color are inactive during the respective read operations.

**Table 1. Benchmarking of CAMs**

|  | [1] | [29] | [30] | [19] | [31] | **This work** |
|---|---|---|---|---|---|---|
| Cell structure | 16T | 2T-2PCM | 2-Flash | 2-FeFET | 6T | 1FeFET-2T |
| Technology (nm) | 45 | 45 | 45 | 45 | 28 | 28 |
| Cell are ($\mu m^2$) | 1.12 | 0.41 | 0.4 | 0.15 | 0.152 | 0.156[a] |
| Speed (ps) | 582 | 155 | 679 | 355 | - | 480[b] |
| Search energy (ft/bit/search) | 1 | 0.64 | 0.6 | 0.4 | 0.6 | 0.3[c] |
| Write energy | 4.8 | 4500 | 98000 | 1.4 | - | 345.6 |
| Search voltage | 1 | 1 | 1.1 | 1 | 1 | 1.5 |
| Storage type | B\|VM | B\|NVM | B\|NVM | T\|NVM | T\|VM | T\|NVM |
| Memory modes (No. of applications per memory block) | 1 | 1 | 1 | 1 | 3 | 3+ |

[a]estimated [24]
[b]Spice simulation of 64 x 64 array
[c]average search energy
B – Binary
T – Ternary
VM – Volatile memory
NVM – Non-Volatile memory

As the ratio $I_{ON}$:$I_{OFF}$ ~30 ($I_{OFF} \approx$ 24 nA), 30 reverse biased diodes will produce equivalent current as one forward biased diode. Therefore, the current difference between the string with the exact match (all diodes reverse biased) and the string with one diode flipped in the exact match string, is $I_{DIFF} = I_{ON} - I_{OFF}$. This is the lowest resolution between two consecutive bit strings, and is independent of the size of the searched strings. However, there are other circuit constraints and non-idealities which put a limit on the maximum length of the search string like the interconnect resistance, the RC delays, and current resolution capability of the sense amplifiers etc. This requires further analysis with the fabricated array, which is beyond the scope of this study.

Table 1 benchmarks the MCAM among the other state-of-the-art and emerging transistor-based CAMs. We observe an area and energy advantage due to lower technology node. From the benchmarking table, we can deduce that this work offers better on some aspects or stands similar to other existing work without significant difference. The NOR structure allows the array to function as a cross-point memory, suitable for the MAC operation along with the analog states of the FeFET (future scope of this work). This level of flexibility makes the memory modes (No. of applications handled by the same memory block) for this work even more than 3.

## 4. CONCLUSION

We have demonstrated the integration of a reconfigurable FeFET with a reconfigurable array and have shown a three-in-one field-programmable memory. The programming of MirrorBit and the MirrorBit-based diode characteristics have been experimentally demonstrated. There is an excellent match between the experimental data and the simulation results. We have also shown the working of the MCAM operation on the reconfigurable structure. The dual flexibility (the device and the array) enables the MirrorBit-integrated FPMA to morph into the application oriented configuration and accomplish the adaptive resource allocation for the energy and performance optimization.

## ACKNOWLEDGEMENTS

This work is supported in part by Department of Science and Technology (DST) Nano Mission, Ministry of Electronics and Information Technology (MeitY), and Department of Electronics, through the Nano-electronics Network for Research and Applications (NNETRA) project of the Government of India and is funded by the German Bundesministerium für Wirtschaft und Klima (BMWK) and by the State of Saxony in the frame of the "Important Project of Common European Interest (IPCEI: WIN-FDSOI)."